%% file: main-cvm-arxiv.tex
\documentclass[10pt,twocolumn,letterpaper]{article}

\usepackage{cvm}
\usepackage{times}
\usepackage{graphicx}
\usepackage{amsmath,amssymb}
\usepackage{booktabs,multirow,tabularx}
\usepackage{xspace}
\usepackage{xcolor}
\usepackage{microtype}

\usepackage[pagebackref=true,breaklinks=true,letterpaper=true,colorlinks,bookmarks=false]{hyperref}

\newcommand{\keywords}[1]{{\bf \emph{Keywords: #1}}}

\cvmfinalcopy 

\def\cvmPaperID{****} 

\ifcvmfinal\fi

\usepackage{amsthm}
\theoremstyle{definition}
\newtheorem{theorem}{Theorem}[section]

\newtheorem{corollary}[theorem]{Corollary}

\newcommand{\score}{\mathcal{G}}
\newcommand{\geoengine}{\mathcal{M}_{\text{geo}}}
\newcommand{\skill}{s}

\newcommand{\harness}{h}

\newcommand{\editbudget}{L_k}

\newcommand{\Acurv}{A_{\mathrm{curv}}}
\newcommand{\Acrease}{A_{\mathrm{crease}}}
\newcommand{\Asig}{n_{\mathrm{sig}}}
\newcommand{\diag}{\mathrm{diag}}


\newcommand{\Description}[1]{}

\begin{document}

\title{RemeshingAgent: An Iterative Agent for Repair-Based Decimation of Generated Building Meshes}

\author{Haipeng Wang\\
Neusoft\\
{\tt\small haipeng\_wang@neusoft.com}
}

\maketitle

\begin{abstract}
Aggressively simplifying a generated high-poly building mesh to $0.5$--$1\%$ of its faces confronts a hard limit. The surface looks complete, yet the mesh is topologically shattered: $10^3$--$10^4$ connected components and non-manifold thin walls. Below a certain budget, every practical decimator ``melts'': it deletes parts, welds thin walls, or breaks the error tolerance. We attribute this to instance-specific pathology: which components are significant and where the cliff sits vary from mesh to mesh, so no single fixed algorithm suffices, and decimation becomes an iterative loop of cleaning, simplifying, and verifying. We present RemeshingAgent, an agent that turns this loop into a validated process. The melting follows from a triangle-budget decomposition into curvature, crease, component, and genus terms: below the component floor, no output satisfies the repair contract, and cleaning first is the theory-mandated response. At a $0.5\%$ budget the agent stays topology-clean where QEM and vertex clustering melt, matches a commercial remesher on 20 unseen city buildings, and on the 24 training buildings a nine-participant single-workstation blind study prefers our outputs $168{:}48$ ($p = 0.0195$, subject-level sign test). The trained skill deploys as retrieval replay: 11.7 seconds per mesh on CPU, with no LLM calls.
\end{abstract}


\input{sections/introduction}   
\input{sections/related}        
\input{sections/theory_method}  
\input{sections/theory1}        
\input{sections/theory3}        
\input{sections/experiments}    
\input{sections/conclusion}     

\bibliographystyle{cvm}
{\small
\bibliography{references}
}

\end{document}

%% file: sections/introduction.tex
\section{Introduction}\label{sec:intro}

\begin{figure*}[!t]
 \centering
 \includegraphics[width=0.76\textwidth]{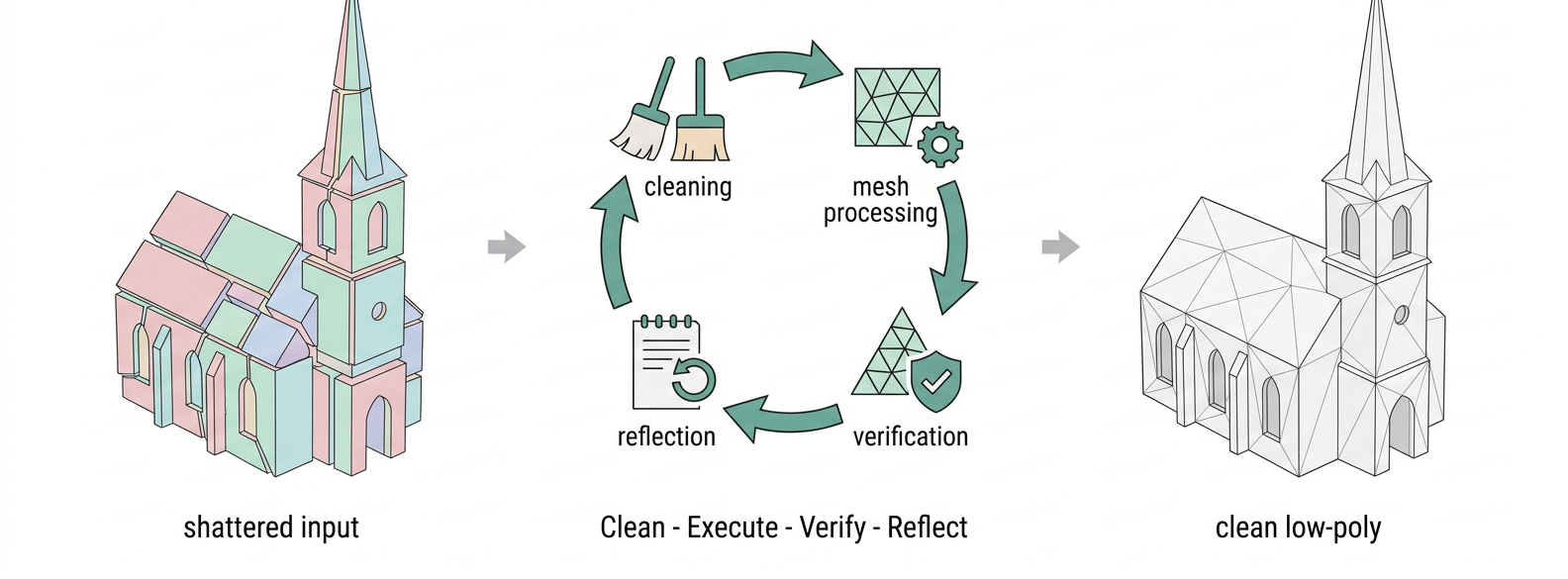}
 \caption{RemeshingAgent at a glance. A generated building looks complete, yet its mesh arrives as shattered fragments (left; Figure~\ref{fig:fragmentation}). The agent's answer is a validated \emph{Clean--Execute--Verify--Reflect} loop (center): clean first, decimate on a frozen engine, verify every candidate, reflect to improve the skill.}
 \label{fig:hero}
\end{figure*}

Mesh decimation is a fundamental operation in interactive rendering, digital twins, and asset pipelines, where high-poly meshes must be reduced to lightweight counterparts. In generated building meshes, a striking discrepancy exists between appearance and structure: the surface appears complete and geometrically sound, yet the underlying mesh is topologically shattered into $10^3$--$10^4$ connected components with abundant non-manifold thin walls (Figure~\ref{fig:hero}, left; Figure~\ref{fig:fragmentation}, left). Because the visible topology is already satisfactory and only the face count is excessive, the natural objective is repair-based decimation. This work addresses the aggressive regime of this problem, namely budgets of $0.5$--$1\%$ of the input faces on generated building meshes.

Existing strategies share a common limitation: all assume that a single fixed policy suffices to reach the target budget. At aggressive budgets this assumption fails—every preservation-contract decimator we test eventually melts on fragmented inputs, manifested as part deletion, wall welding, or tolerance violation. Existing theory neither predicts where melting must occur—its budget results cover smooth convex bodies~\cite{fejes1948approximation,glasauer1996asymptotic} and plane curves~\cite{fejes1948approximation} but not fragmented piecewise-smooth surfaces—nor addresses whether a self-improving decimator converges or how much validation evidence suffices to certify improvement. The absence of theoretical guidance is consequential: each mesh shatters differently, rendering run-once policies inadequate; manual orchestration of the cleaning-simplification-verification loop is prohibitively tedious; and unguided agent-based approaches are unstable, as erroneous early decisions propagate irrecoverably. Furthermore, once the decimation process is governed by a strict validation gate, it becomes a drift process whose convergence can be quantified.

Our central insight is that the melting point is not an algorithmic artifact but a geometric property of the input itself, and therefore a theory-mandated strategy can be derived from measured geometry rather than discovered through blind search. We therefore present RemeshingAgent, an agent for repair-based decimation of generated building meshes that combines a frozen geometry engine with a validation-gated evolution discipline~\cite{luo2026skillopt}. The agent operationalizes the decimation craft as a validated \emph{Clean--Execute--Verify--Reflect} loop (Figure~\ref{fig:hero}, center--right): it cleans the fragmented input, executes decimation pipelines on the frozen engine, verifies candidates against held-out geometric scores and visual inspection, and reflects on the trajectory to refine its skill document and operator vocabulary, accepting modifications only when they strictly improve held-out validation. 

Empirically, trained on 24 commercial high-poly building meshes with 19 held-out validation meshes, the agent improves its gate score monotonically from $8.8358$ to $8.9811$ across five accepted updates; deployed inference replays the retrieved pipelines on CPU in $11.7$\,s per mesh without LLM or VLM calls.

\subsection{Contributions}

\begin{itemize}
 \item We present RemeshingAgent, the first agent for repair-based decimation of generated building meshes, turning fragment clouds of thousands of components into single clean components (Figure~\ref{fig:fragmentation}, right).
 \item We give the agent a self-evolving discipline that tunes the skill document and the operator vocabulary along the theory's direction chain under a strict held-out validation gate, improving by $+0.145$ gate score without a single regressed update.
 \item We prove the theory behind both, a budget decomposition prescribing the clean-first direction, and a component threshold below which no output satisfies the repair contract.
\end{itemize}

First, the triangle budget admits a decomposition into a curvature term $\Acurv/\varepsilon$, a crease term $\Acrease/\sqrt{\varepsilon}$, and a component floor $4\Asig$, with the maximum determining the active constraint (Theorem~\ref{thm:budget}). Second, below the component threshold, no output satisfying the repair contract exists, transforming melting from an empirical observation into a prediction with a precise location, and establishing cleaning as the theory-mandated response (Theorem~\ref{thm:cliff}, Corollary~\ref{cor:clean}). Third, under an explicit improvement assumption, the gated evolution admits an additive-drift hitting-time bound together with a resolution law characterizing the validation-set size required to resolve improvements (He--Yao~\cite{he2001drift}, Eq.~\eqref{eq:resolution}), with all parameters estimated from the agent's training trace.

%% file: sections/related.tex
\section{Related Work}\label{sec:related}

\textbf{LLM Agents for 3D.} LLM-based agents have been applied to 3D generation~\cite{poole2022dreamfusion} and procedural modeling. ReAct~\cite{yao2023react} and Voyager~\cite{wang2023voyager} establish observation-action loops; PhotoAgent~\cite{photoagent2026}, aDSL~\cite{wang2026adsl}, and CADDesigner~\cite{fan2026caddesigner} couple LLM planning with graphics tools—none evolves its own operator vocabulary under validation discipline, establishes convergence guarantees, or addresses repair-based decimation at aggressive budgets. Our approach treats the entire remeshing policy as a trainable object, with convergence analyzed in Section~\ref{sec:theory3}.

\textbf{Level of Detail and Decimation.} Classical approximation theory characterizes the triangulation complexity of smooth surfaces: the $\Theta(\int\sqrt{|K|}\,dA/\varepsilon)$ face-count bound for convex surfaces~\cite{fejes1948approximation,glasauer1996asymptotic}, Fejes T\'oth's optimal polygonal curves~\cite{fejes1948approximation}, and Erickson's sampling complexity~\cite{erickson2003nice}---but none of these addresses the repair-contract minimum face count of fragmented piecewise-smooth surfaces; field reports~\cite{li2024simplifying} document such failures without a theoretical account. Preservation-based decimators maintain input connectivity (quadric error metrics~\cite{garland1997surface}, progressive meshes~\cite{hoppe1996progressive}, vertex clustering, recent variants for weak features and artist quad topology~\cite{faqem2026,knodt2024quad}; surveys~\cite{cignoni1998comparison,botsch2010polygon,luebke2003level}), whereas reconstruction rebuilds the surface from samples (voxel hulls, alpha wrap, cell cuts~\cite{yang2026cellocut}, generative remeshers~\cite{li2026lato2}). City-scale LOD frameworks~\cite{biljecki2015applications} prescribe level contents but not decimation limits; our budget law supplies this missing quantitative layer. Head-to-head comparisons with these robust baselines appear in the supplementary (S11--S12). Our two-contract analysis (supplementary, S10) predicts when each family prevails; learned representations~\cite{mescheder2019occupancy,park2019deepsdf,hanocka2019meshcnn,li2024advances3d} need training sets incompatible with our single-mesh setting.

\textbf{Self-Evolving Agents.} Self-improving frameworks (Reflexion~\cite{shinn2024reflexion}, G\"odel Agent~\cite{yin2025godel}) update task and meta policies; prompt optimizers (DSPy~\cite{khattab2023dspy}, TextGrad~\cite{yuksekgonul2024textgrad}, GEPA~\cite{agrawal2025gepa}) refine instructions. SkillOpt~\cite{luo2026skillopt} contributes the discipline we adopt—rollout batches, bounded edits, validation gating, a rejected-edit buffer—for a frozen geometry engine. RE-TRAC~\cite{retrac2026} addresses cross-trajectory state compression, and complementary work distills skills from multimodal resources~\cite{resource2skill2026} and evaluates them via checklists~\cite{bizgeneval2026}. E-value gating~\cite{sgm2025statistical,lee2026sea} and validation-gated skill evolution~\cite{valve2026} provide error-control guarantees but leave open the power analysis behind our resolution law (Eq.~\eqref{eq:resolution}) and our planner's multi-candidate selection; our convergence analysis builds on the additive-drift theorem~\cite{he2001drift} and hitting-time bounds~\cite{akimoto2018theoretical}.

%% file: sections/theory_method.tex
\section{The Geometry of Decimation}\label{sec:theory}

Section~\ref{sec:theory1} derives the agent's direction from measured geometry, and Section~\ref{sec:theory3} adds a self-evolving loop that verifies every candidate against held-out scores.

\begin{figure}[t]
    \centering
    \includegraphics[width=0.82\columnwidth]{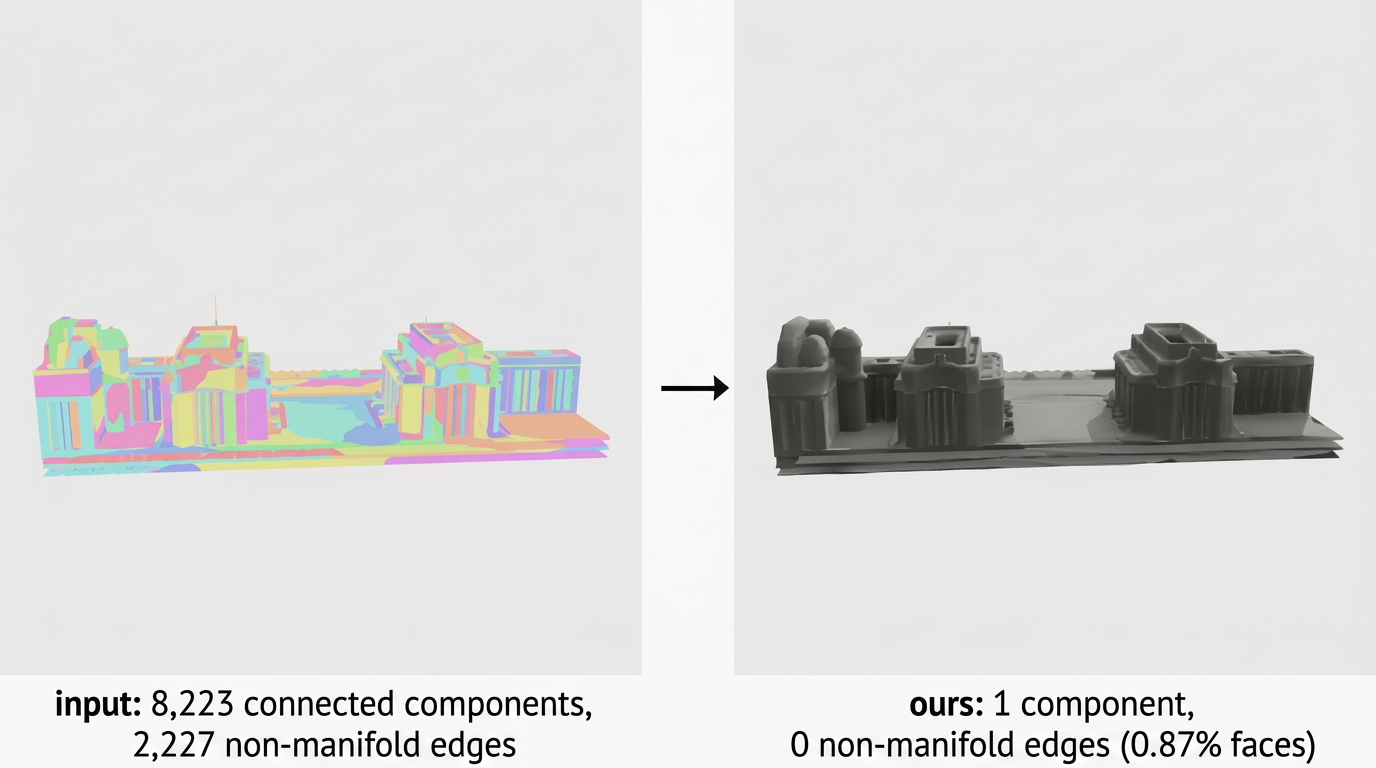}
    \caption{A representative generated building: complete-looking surface, shattered interior—$8{,}223$ components, $2{,}227$ non-manifold edges (left, components randomly colored). Our output: one component, zero non-manifold edges, $0.87\%$ of the input faces (right).}
    \label{fig:fragmentation}
\end{figure}

%% file: sections/theory1.tex
\subsection{The Minimal Triangle Budget}\label{sec:theory1}

A generated building's mesh arrives shattered into $10^3$--$10^4$ components (Figure~\ref{fig:fragmentation}, left); the objective is repair-based decimation—consolidating $8{,}223$ components into one at $0.87\%$ of the input faces (right). 

\subsubsection{The Topology-Preserving Contract}\label{sec:contract}

We model the input $S$ as piecewise-smooth with $n$ components of total genus $g$; a mesh $T$ is an $\varepsilon$-approximation if $\delta_H(S,T) \le \varepsilon\,\diag(S)$. Topology is measured against the \emph{intended surface} $S^\ast$, the welding closure of $S$ at the seam scale $\delta_0$: the intended building parts, pairwise separated well beyond $\varepsilon\,\diag(S)$ at our budgets. $S^\ast$ is a \emph{declared} reference computed from $S$ alone—generated assets carry no pre-fragmentation ground truth—so preservation is always relative to $S^\ast$: welding and dust removal repair $S$ toward $S^\ast$ by definition; Section~\ref{sec:exp-testset} audits this structure component-wise. Under the \emph{preservation contract} (A), $T$ must satisfy the tolerance and $T \cong S^\ast$ (homeomorphic)—welding is repair, not a topology change (raw-input preservation would force $F \ge 4n$ if every component were a closed shell—a floor no budget meets). In the agent the welding threshold defaults to this seam scale, $0.1\%$ of the diagonal, and adopted pipelines kept it throughout. Under the \emph{reconstruction contract} (B), the output is rebuilt from samples (voxel hulls, alpha wrap, generative remeshers~\cite{li2026lato2}); its budget law differs. A component is \emph{$\varepsilon$-significant} if its diameter exceeds $2\varepsilon\,\diag(S)$; deleting it violates the approximation. Significant components are closed shells (thin-walled solids); boundary components admit lower floors (supplementary, S10).

\subsubsection{The Budget Law}\label{sec:building}

\begin{theorem}[Minimal triangle budget, decomposition]\label{thm:budget}
Under the preservation contract, as the absolute tolerance $\varepsilon_{\mathrm{abs}} = \varepsilon\,\diag(S) \to 0$,
\begin{equation}\label{eq:budget}
F_{\min} \;\asymp\; \max\!\big( c_1 \Acurv/\varepsilon_{\mathrm{abs}},\, c_2 \Acrease/\sqrt{\varepsilon_{\mathrm{abs}}},\, 4\,\Asig,\, c_4 g \big),
\end{equation}
where $\Acurv = \int_{S_{\mathrm{smooth}}}\sqrt{|K|}\,dA$ and $\Acrease = \sum_j\int_{C_j}\sqrt{\kappa}\,ds$, and $c_1,c_2,c_4$ depend only on the smoothness constants; $\asymp$ is two-sided asymptotic equivalence.
\end{theorem}

Each term is classical~\cite{fejes1948approximation,glasauer1996asymptotic,ludwig1998asymptotic,jungerman1980minimal}. Developable patches ($K = 0$)—cylindrical towers—cost $\Theta(\varepsilon_{\mathrm{abs}}^{-1/2})$ and fold into the crease term (prism bound $\pi\sqrt{R/(2\varepsilon_{\mathrm{abs}})}$); the distinct exponents ($\varepsilon_{\mathrm{abs}}^{-1}$, $\varepsilon_{\mathrm{abs}}^{-1/2}$, $\varepsilon_{\mathrm{abs}}^0$) make the decomposition identifiable. For predominantly planar, low-genus buildings, Eq.~\eqref{eq:budget} reduces to
\begin{equation}\label{eq:building}
F_{\min} \;\asymp\; \max\!\Big( c_2\,\frac{\Acrease}{\sqrt{\varepsilon_{\mathrm{abs}}}},\; 4\,\Asig \Big),
\end{equation}
so classical analysis alone predicts near-free decimation—and the missed component floor $4\Asig$ is what makes every practical decimator we test melt on generated ones; the operational reading—clean first, weld, then decimate—follows (Figure~\ref{fig:fragmentation}, right).

Theorem~\ref{thm:budget} is asymptotic; at $0.5$--$1\%$ budgets we claim only the \emph{regional structure} as evidence, validated in Section~\ref{sec:exp-regime}.

\subsubsection{The Operational Limit}\label{sec:estimator}

Fitted per algorithm family with a calibration factor $\hat c$,
\begin{equation}\label{eq:estimator}
\hat F(S,\varepsilon) = \hat c_1 \frac{\Acurv}{\varepsilon\,\diag} + \hat c_2 \frac{\Acrease}{\sqrt{\varepsilon\,\diag}} + 4 \Asig,
\end{equation}
Equations~\eqref{eq:budget}, \eqref{eq:building}, and \eqref{eq:cliff-threshold} use the absolute tolerance $\varepsilon_{\mathrm{abs}}$ defined above; $\varepsilon$ elsewhere is the contract's dimensionless relative tolerance, so every term is a face count. $\Asig$ is measured on the post-cleaning, welded estimate of $S^\ast$ (the surface the contract references), with the $0.1\%$ area threshold (rank-consistent with the diameter caliber, $\rho = 0.68$--$0.89$). Fitting QEM and clustering sweeps gives $\hat c \in [1.1\times, 2.4\times]$ (Section~\ref{sec:exp-regime}): an additive working form, its max-to-sum gap absorbed by $\hat c$—calibrated on the sweep corpus, not predicted beyond it; the tested content is the exponent structure and the cliff shift, and a split-half probe predicts held-out error floors rank-consistently (Spearman $\rho = 0.83$ median; supplementary, S6).

\begin{theorem}[Component cliff]\label{thm:cliff}
Let $S^\ast$ have $\Asig$ $\varepsilon$-significant components, pairwise separated by more than $2\varepsilon\,\diag(S)$—true of intended building parts at our budgets—and take a budget
\begin{equation}\label{eq:cliff-threshold}
F < F_{\mathrm{comp}} \;:=\; \sum_{k=1}^{\Asig} \max\!\big(4,\, c_1 \Acurv(S_k)/\varepsilon_{\mathrm{abs}}\big).
\end{equation}
Then no mesh $T$ with at most $F$ triangles satisfies both $\delta_H(S^\ast,T) \le \varepsilon\,\diag(S)$ and $T \cong S^\ast$: the separation keeps the components' $\varepsilon\,\diag(S)$-neighborhoods disjoint, so no triangle serves two components, and the piece of $T$ standing for each closed shell $S_j$ pays its own floor,
\begin{equation}\label{eq:cliff-error}
F \;\ge\; \sum_{k=1}^{\Asig} \max\!\big(4,\, c_1 \Acurv(S_k)/\varepsilon_{\mathrm{abs}}\big) \;=\; F_{\mathrm{comp}} \;>\; F,
\end{equation}
\end{theorem}

\begin{figure*}[t]
    \centering
    \includegraphics[width=0.56\textwidth]{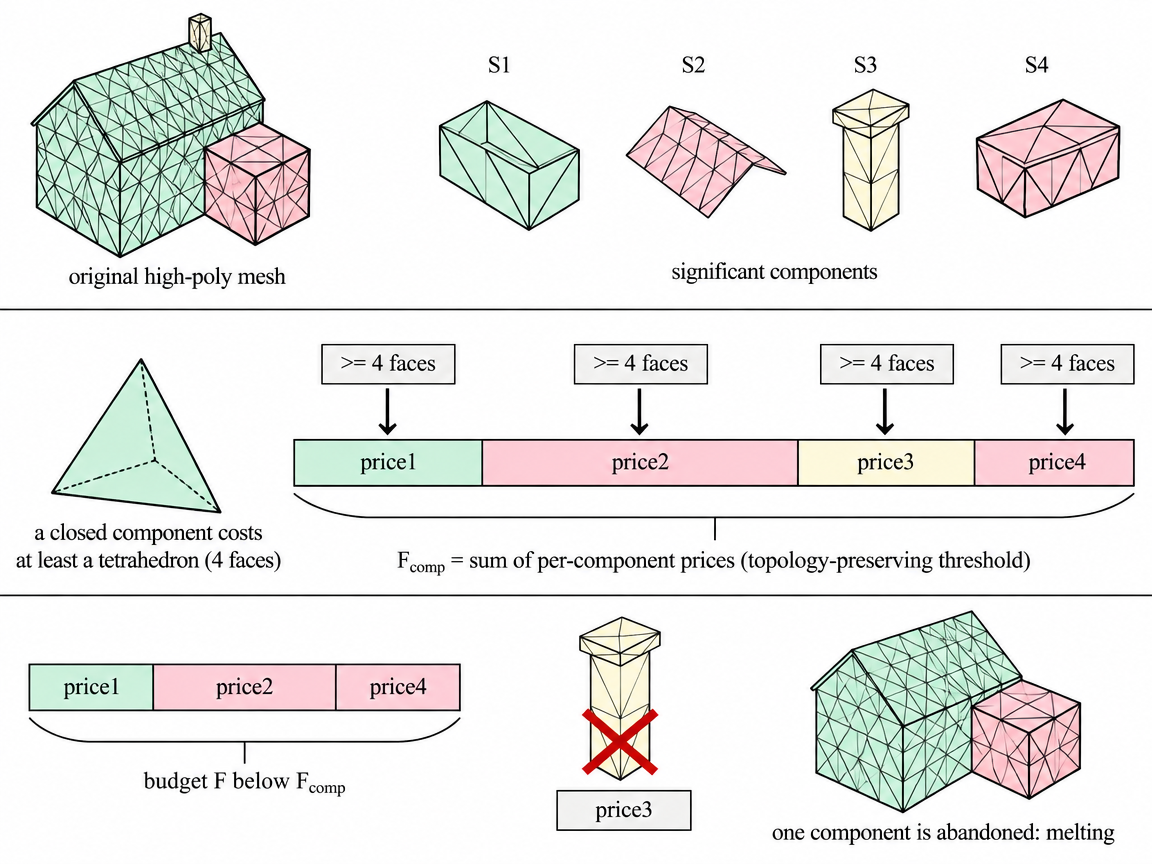}
    \caption{The component cliff as a counting argument. The high-poly input decomposes into significant components (S1--S4), each costing at least four dedicated faces; their prices sum to the threshold $F_{\mathrm{comp}}$ of Theorem~\ref{thm:cliff}, and a budget below it abandons a component entirely (here the chimney).}
    \label{fig:cliff}
\end{figure*}

The bound is \emph{combinatorial}, not statistical; it holds for every algorithm (Figure~\ref{fig:cliff}; proof and breaking-mode classification in the supplementary, S10).

\begin{corollary}[The value of cleaning]\label{cor:clean}
A component is \emph{$\varepsilon$-deletable} if every point of it lies within $\varepsilon\,\diag(S)$ of the retained surface; deleting such components preserves the $\varepsilon$-approximation. Dust is deletable by construction—shards of the intended shell—so cleaning lowers $\Asig$ from $n$ to $\Asig \ll n$ and $F_{\mathrm{comp}}$ accordingly. Cleaning before decimation is the theory-mandated first move (supplementary, S7).
\end{corollary}

\emph{From fragments to one component.} Three theory-licensed moves: delete dust (Corollary~\ref{cor:clean}), weld vertices closer than $\delta \le \varepsilon\,\diag$ (perturbs by at most $\delta$), then pay the budget law at the reduced $\Asig$—realized by the agent's \emph{clean\_repair} (conservative cleanup, or aggressive welding plus small-component removal), stopping before the threshold rather than melting (Section~\ref{sec:exp-zoo}); the operators' own perturbation is $0.0009$--$0.0017$ normalized Chamfer, within tolerance.

%% file: sections/theory3.tex
\subsection{Convergence of Validation-Gated Skill Evolution}\label{sec:theory3}

\begin{figure}[t]
    \centering
    \includegraphics[width=0.68\columnwidth]{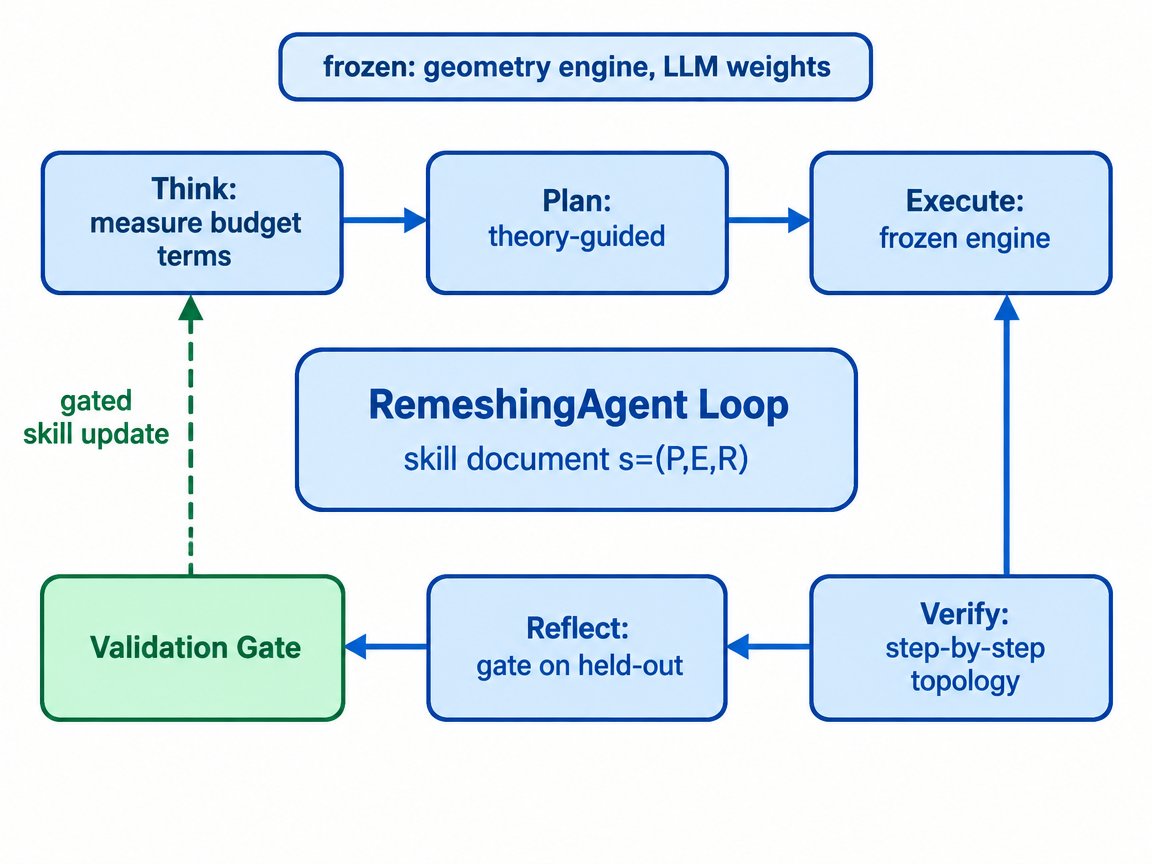}
    \Description{Diagram of the RemeshingAgent loop: Think, Plan, Execute, Verify, Reflect, with the validation gate and the skill document as the core trainable state.}
    \caption{RemeshingAgent loop. The skill document $s=(P,E,R)$ is the core trainable state; retrieval weights and planner prompt evolve under the same gate; the geometry engine and the LLM are frozen.}
    \label{fig:system}
\end{figure}

Figure~\ref{fig:system} illustrates the loop: the direction of Section~\ref{sec:theory1} fixes what to do but not its parameters, which the agent tunes by evolving a skill document $s$ through reflect-and-gate cycles. With selection set $D_{\mathrm{sel}}$ and deterministic reward $r(s,m)$, the \emph{empirical potential} is $V(s) = \mathrm{target} - \frac{1}{|D_{\mathrm{sel}}|}\sum_{m} r(s,m)$, with hitting time $T_\varepsilon = \min\{t : V(s_t) \le \varepsilon\}$. If the reflect kernel $Q(s \to s')$ satisfies condition (H1)—improving $V$ by at least $\Delta$ with probability $\ge p_{\min}$ (supplementary, S10)—and the gate uses strict acceptance with elite retention, the additive-drift bound of He--Yao~\cite{he2001drift} yields $\mathbb{E}[\,T_\varepsilon \mid s_0\,] \le V(s_0)/(p_{\min}\Delta)$; calibrated on our system, it agrees in order of magnitude with the observed trajectory—a consistency check, not an independent prediction. The per-group law $p\Delta$ is the testable content.

\subsubsection{The Validation Gate}\label{sec:tier1}

The reward is a scalarized score $\score(M; M_0) = \sum_{i=1}^{6} w_i\, g_i$ with $w = (0.30, 0.15, 0.20, 0.05, 0.20, 0.10)$ over face-count, manifold, shape-preservation, planarity, sharp-edge, and compactness dimensions (calibrated in Section~\ref{sec:exp-evolution}; supplementary S9). The gate's statistical power limits the smallest per-step improvement it can resolve—the sub-Gaussian resolution scale
\begin{equation}\label{eq:resolution}
\Delta_{\min}(n) \;=\; \hat\sigma \sqrt{\frac{2\ln(1/\alpha)}{n}},
\end{equation}
with $\hat\sigma$ the cross-mesh std of paired differences.

\subsubsection{Forward and Backward Passes}

The harness returns scored trajectories $(\tau, r) = \harness(\geoengine, M_0, \skill)$; at round $k$ the planner samples $B$ candidates and submodular selection picks a diverse subset. A pipeline template in $\skill$ is the parameterized chain of Section~\ref{sec:theory1} made executable. An LLM \emph{reflector} proposes discrete text edits over the success/failure split, guided by a cross-trial \emph{parameter$\to$reward contrast table}; accepted edits are capped by $\editbudget$ (the ``learning rate''), yielding candidate $\tilde\skill_k$, gated on the held-out set:
\begin{equation}\label{eq:gate}
\skill_{k+1} = \tilde\skill_k \iff r_{\mathrm{sel}}(\tilde\skill_k) > r_{\mathrm{sel}}(\skill_k);
\end{equation}
otherwise $\skill_{k+1} = \skill_k$ (elite retention); a pre-check skips ``inert'' candidates. Candidates run in an isolated read-only copy, so rejections never enter the deployed skill; the best score is non-decreasing and the initial skill is a safe lower bound (supplementary, S10). Because $D_{\mathrm{sel}}$ is reused, the deployed skill is re-scored on a test set it never touches (Section~\ref{sec:exp-testset}): test numbers measure transfer, not gate fitness.

\subsubsection{The Rejected-Edit Buffer and the Tool Tier}\label{sec:tier2}

Rejected edits are retained as negative feedback; at epoch boundaries a meta-reflector writes longitudinal guidance into a protected slow-update region. Skill training optimizes \emph{what the agent knows} but cannot escape its operator vocabulary; the tool tier removes that ceiling—an LLM proposes \emph{new operators}, each passing a feasibility probe, with adoption adjudicated by the ordinary gate (Section~\ref{sec:exp-operator})—and no gate is ever relaxed.

%% file: sections/experiments.tex
\section{Experiments}\label{sec:experiments}

\subsection{Datasets and Baselines}

The training set comprises 24 high-poly Hunyuan3D building meshes (0.9M--1.5M faces, 7{,}000--21{,}000 components), each decimated to three budgets (1\%, 0.75\%, 0.5\%); the validation set holds 19 landmark meshes ($\sim$740K faces) for gate evaluation; and the theory corpus covers the same 43 meshes, 12 of which also run QEM/clustering sweeps at eight budgets (10\%--0.1\%). We evaluate on \emph{every} mesh in each split: failed epochs are scored, all 19 validation meshes enter every gate round, no mesh is added, removed, or re-weighted after results, and group order follows a pre-registered seed.

Five method families are compared. Direct QEM is quadric-error decimation (pymeshlab, \emph{preserve-topology} enabled); vertex clustering is \emph{meshing\_decimation\_clustering}; the zoo additionally reports QEM with \emph{preserve-topology} disabled (the only configuration reaching aggressive budgets). LATO.2~\cite{li2026lato2} is a learned generative remesher served on its released inference server with the vertex budget specified directly; our pipeline and all preservation baselines are CPU-only. The reconstruction pipeline chains voxel shell $\to$ nvdiff deformation $\to$ alpha wrap $\to$ refined QEM as the contract-B representative. RemeshingAgent runs zero-shot ($S_0$) and trained ($S_K$, $K=24$ groups).

All experiments run on CPU (pymeshlab, trimesh, Open3D); the LLM planner and reflector employ MiniMax-M3, and the VLM bonus is disabled in training for deterministic evaluation. Reward weights derive from a calibration episode (an initial planarity weight admitted degenerate slivers); recalibration (shape/sharp reweighting, a non-linear cap $>7.0$, a degenerate-face penalty) aligned gate decisions with perceptual quality, robust under $\pm 50\%$ and $100$ in-band perturbations (supplementary, S9).

\subsection{Self-Evolution of the Agent}\label{sec:exp-evolution}

\begin{figure}[t]
\centering
\includegraphics[width=0.72\columnwidth]{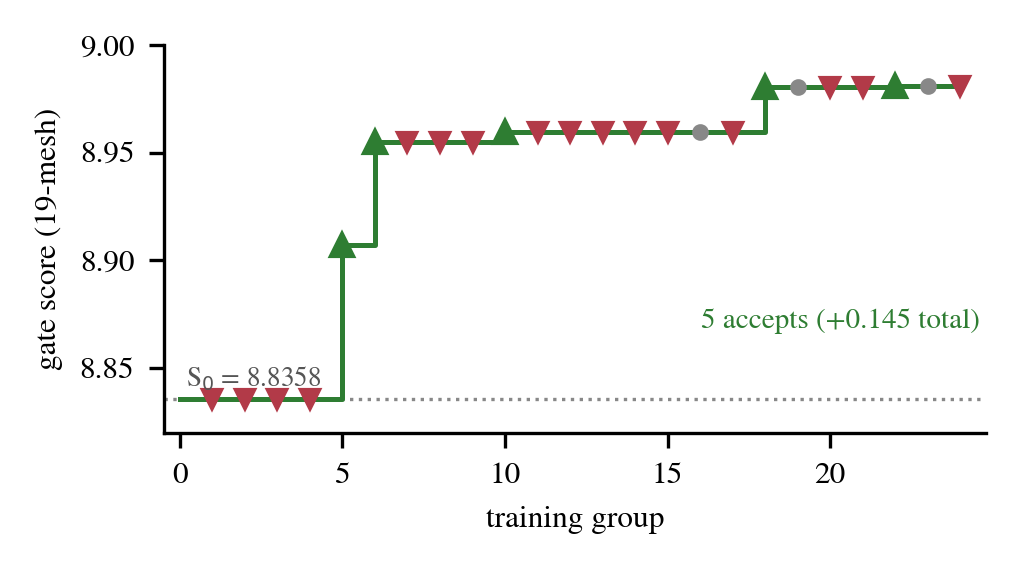}
\Description{Validation gate score over 24 training groups; accepted updates marked.}
\caption{Self-evolution gate trajectory over 24 training groups (19-mesh evaluation; $S_0{=}8.8358 \to 8.9811$). Green triangles: accepted updates; red inverted triangles: rejected.}
\label{fig:rounds}
\end{figure}

Figure~\ref{fig:rounds} shows the gate score over 24 groups: $S_0 = 8.8358 \to 8.9811$ ($+0.145$) through 5 accepted updates; all 19 validation meshes improve (supplementary, S1b), the smallest gain $+0.021$. The drift law predicts $24\,p\Delta \approx 0.115$ vs.\ observed $+0.145$.

\subsection{The Two-Regime Structure}\label{sec:exp-regime}

\begin{figure}[t]
    \centering
    \includegraphics[width=0.70\columnwidth]{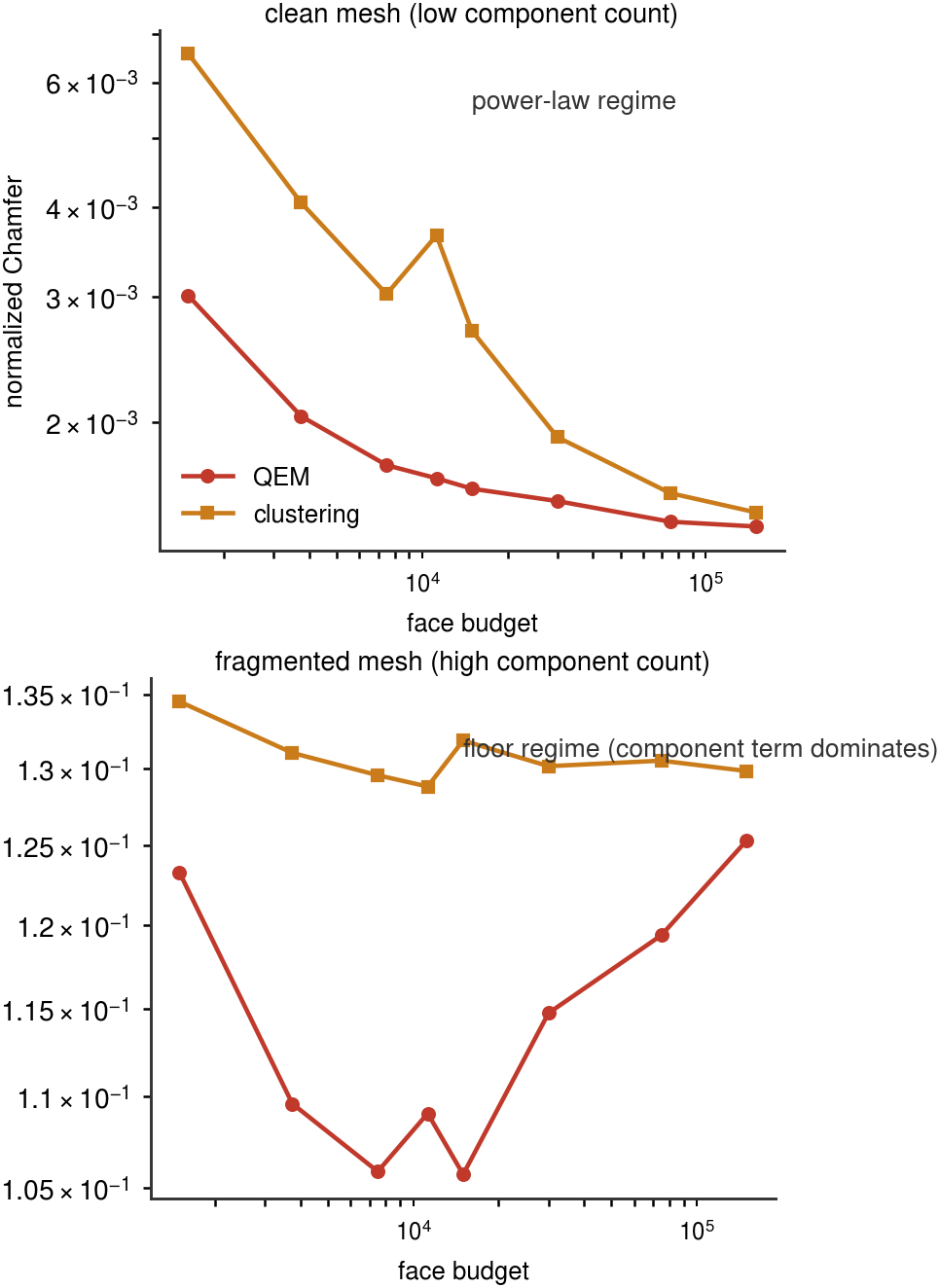}
    \caption{Two-regime structure of error-versus-budget curves. Normalized mean Chamfer error versus budget (log-log) for the QEM and clustering sweeps: power-law regime on clean meshes, floor regime on fragmented meshes.}
    \label{fig:tworegime}
\end{figure}

QEM and clustering sweeps (12 meshes $\times$ 8 budgets) exhibit a two-regime structure (Figure~\ref{fig:tworegime}). On clean meshes, error follows a power law with exponent $0.13$--$0.40$ ($\mathrm{corr}(\log A, \log \mathrm{budget}) = 0.95$); on fragmented meshes, error exhibits a floor regime with exponent $\approx 0$ for the melting families, decreasing only until the component term $4\Asig$ of Theorem~\ref{thm:budget} becomes dominant. The cliff position shifts with the predicted $F_{\mathrm{comp}}$, consistent with component-dominated breaking. Both algorithm families exhibit identical behavior, indicating that the regional structure is a geometric property rather than an algorithmic artifact.

\subsection{The Five-Family Zoo}\label{sec:exp-zoo}

\begin{figure}[t]
    \centering
    \includegraphics[width=0.70\columnwidth]{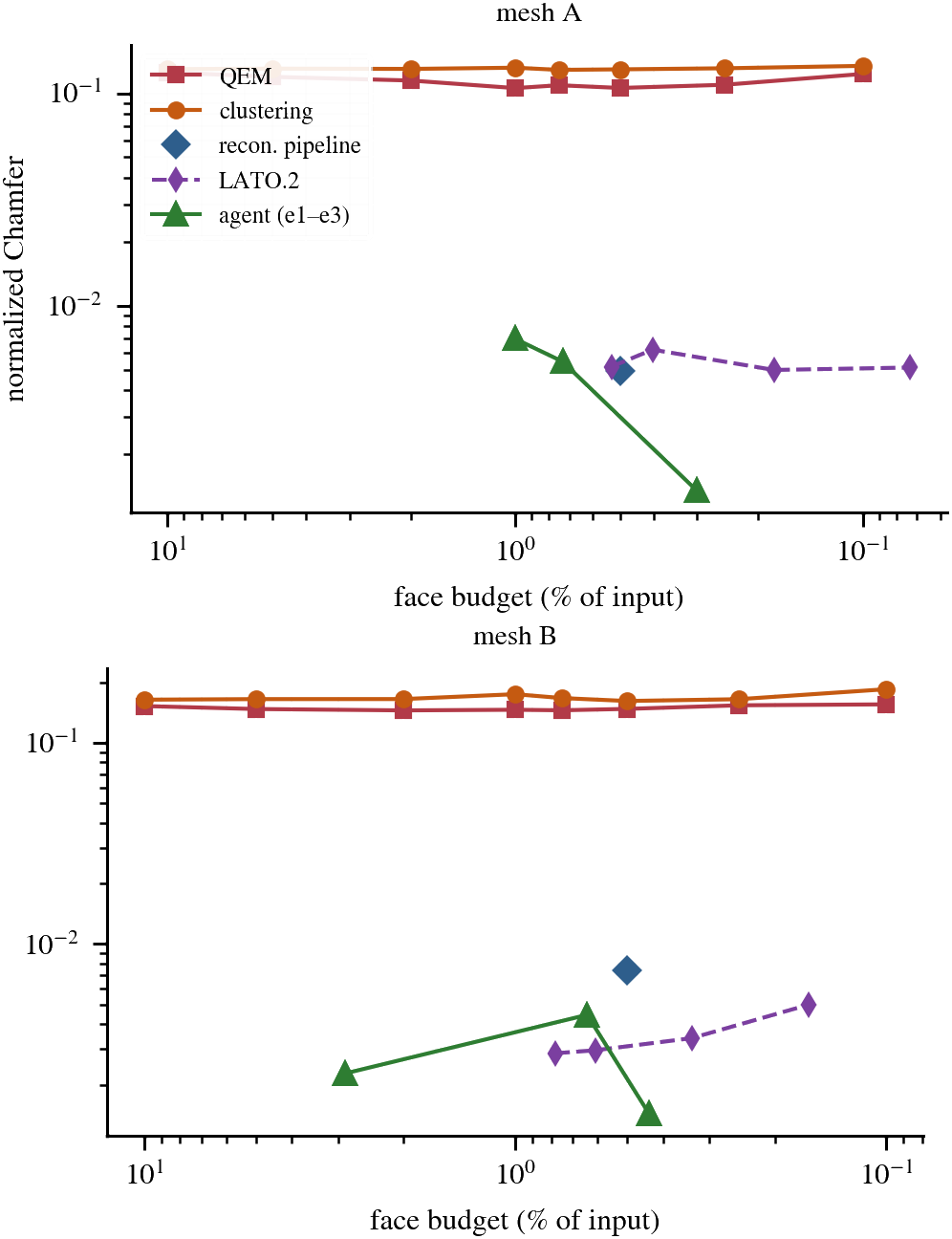}
    \caption{The zoo as per-mesh curves: normalized Chamfer versus face budget on two representative fragmented training meshes (log-log). QEM (red) and clustering (orange) stay flat at the melting level at every budget, while the reconstruction pipeline (blue diamonds) and the agent (green triangles) sit one to two orders of magnitude below, with LATO.2 (purple) in between.}
    \label{fig:zoocurves}
\end{figure}

All five families target the same $0.5\%$ budget on identical fragmented meshes, measured by normalized Chamfer (50k area-weighted samples, symmetric bidirectional mean, normalized by the output diagonal). Reached face counts differ and are part of the comparison: preservation-enabled QEM cannot reach the target, whereas melting configurations reach it by deleting and welding. The agent achieves the best fidelity at $0.0015$ (24/24); the reconstruction pipeline $0.005$--$0.009$ (watertight); LATO.2 $0.003$--$0.011$, floored at low budgets by its generative prior; bare QEM and clustering $0.05$--$0.18$, melting. An equal-cleaning control separates recipe from adaptation: conservative clean-then-QEM reaches the budget on $1/24$, the fixed aggressive recipe on $19/24$ ($0.0014$ median Chamfer on hits, clean topology)—cleaning carries most of the effect; the rest is per-mesh variation fixed parameters miss and selection covers (the agent: $24/24$; supplementary, S7). The topology metrics and Section~\ref{sec:exp-frag} carry the preservation verdict. Figure~\ref{fig:zoocurves} shows the same separation per mesh.

\begin{table}[t]
\centering\footnotesize
\setlength{\tabcolsep}{4pt}
\caption{Regime-conditional necessity: budget-and-topology success under the two distributions at matched relative budgets. Success = faces within the budget band and zero non-manifold edges. On the shattered training distribution the method is necessary (24/24 vs.\ 19/24 vs.\ 1/24); on the simpler held-out buildings commodity tools complete---and aggressive cleaning \emph{induces} its own failure, stalling QEM on 2/20 meshes (760k/660k faces remaining); per-mesh selection avoids both failure modes.}
\label{tab:regime}
\begin{tabular}{lrr}
\toprule
 & \textbf{Hunyuan train (24)} & \textbf{DEU test (20)} \\
\midrule
bare QEM (preserve-topology) & 1/24 & 20/20 \\
conservative clean + QEM & 1/24 & 20/20 \\
aggressive clean + QEM (fixed) & 19/24 & 18/20 \\
adaptive rule (stats-only, no LLM) & -- & 18/20 \\
\textbf{RemeshingAgent (deployed)} & \textbf{24/24} & \textbf{20/20} \\
\bottomrule
\end{tabular}
\end{table}

A regime-conditional stress test makes the scope explicit (Table~\ref{tab:regime}): on Hunyuan-style shattered inputs the method is necessary, while on the independently-sourced test buildings every arm completes at the same relative budgets---and aggressive cleaning \emph{induces} its own melt, stalling QEM on 2/20 meshes with 760k/660k faces remaining; per-mesh selection avoids both failure modes. The Think-stage fragmentation statistics identify the regime before any run.

The ordering matches the two-contract prediction (supplementary, S10): preservation methods pay the component tax or melt, and the agent's cleaning reduces $\Asig$ to the preservation side. The epoch-3 cliff-avoidance response (face-count dimension $-4$ at $0.5\%$ budget, Chamfer holding at $0.0015$) and PolyFit's contract-B verdict ($18/20$ test meshes complete, immaculate topology, median Chamfer $0.029$ vs.\ ours $0.001$; supplementary, S7) complete the picture.

\subsection{Isolating the Component Term}\label{sec:exp-frag}
\begin{figure}[t]
    \centering
    \includegraphics[width=0.72\columnwidth]{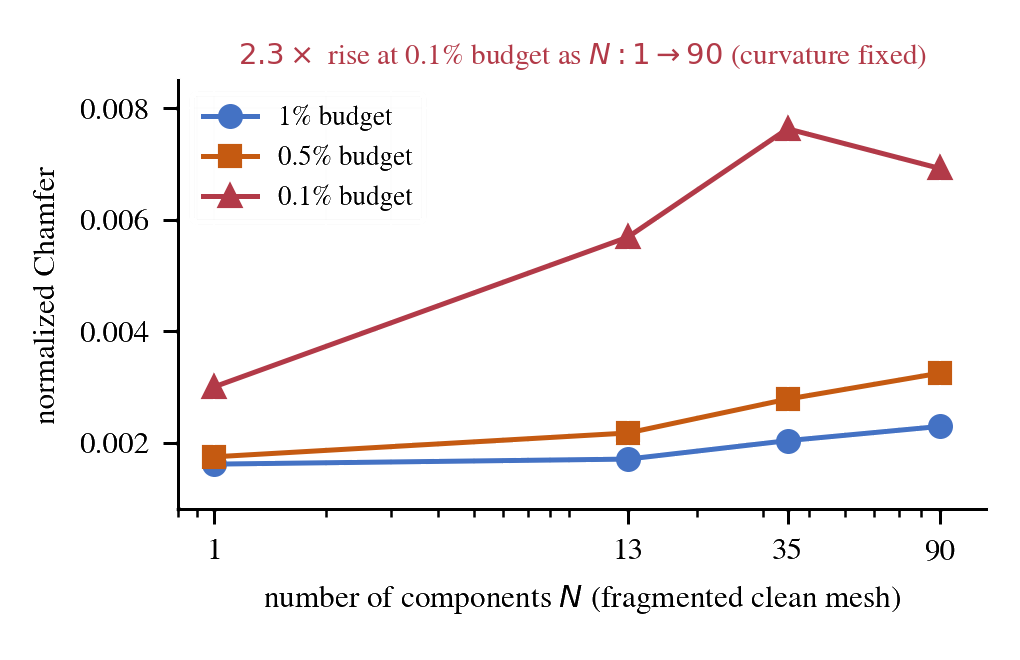}
    \caption{Controlled fragmentation benchmark: normalized Chamfer versus component count $N$ at three budgets on a constant-curvature mesh—the $0.1\%$ curve rises $2.3\times$ from $N{=}1$ to $N{=}90$ with curvature held fixed.}
    \label{fig:fragbench}
\end{figure}

Controlled fragmentation is applied to a clean constant-curvature mesh ($\int\sqrt{|K|}dA \approx 26.2$; $N = 1, 13, 35, 90$ pieces). At the most aggressive budget ($0.1\%$), normalized Chamfer error rises $0.0030\to0.0069$ ($2.3\times$) while curvature stays fixed. The component term is independent and identifiable (Figure~\ref{fig:fragbench}; supplementary S6). Against known truth the lesson repeats: shattering 19 clean landmarks into hundreds of pieces, the fixed clean-then-QEM recipe recovers the true component count to a median error of $1.5$ parts on $\le 20$-part buildings but loses $16$ on more complex ones, while preservation-QEM leaves component clouds in the hundreds (supplementary, S6). 

\begin{figure}[t]
    \centering
    \includegraphics[width=0.64\columnwidth]{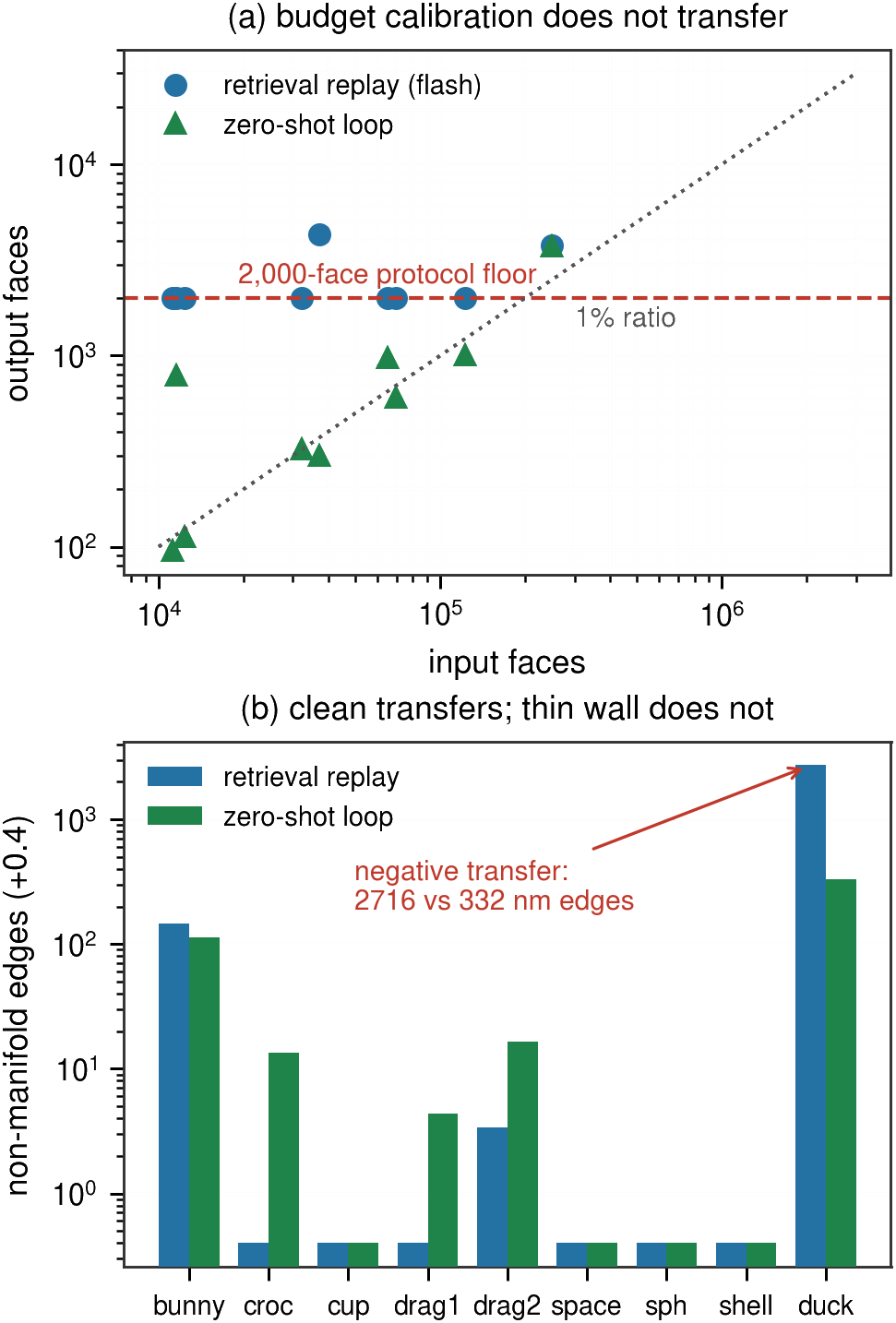}
    \caption{Cross-domain transfer on nine non-architectural meshes.}
    \label{fig:transfer}
\end{figure}

\begin{table}[t]
\centering
\small
\caption{Cross-domain transfer, per mesh: input faces; retrieval-replay output faces and non-manifold (nm) edges; zero-shot-loop output faces and nm edges.}
\label{tab:transfer}
\begin{tabular}{lrrrrr}
\toprule
 & & \multicolumn{2}{c}{\textbf{retrieval replay}} & \multicolumn{2}{c}{\textbf{zero-shot loop}} \\
\cmidrule(lr){3-4}\cmidrule(lr){5-6}
\textbf{mesh} & \textbf{in faces} & faces & nm & faces & nm \\
\midrule
bunny & 69{,}451 & 1{,}999 & 145 & 600 & 112 \\
crocodile & 11{,}222 & 2{,}000 & 0 & 95 & 13 \\
cup & 122{,}778 & 2{,}000 & 0 & 1{,}000 & 0 \\
dragon & 249{,}882 & 3{,}784 & 0 & 3{,}740 & 4 \\
dragon2 & 11{,}518 & 1{,}999 & 3 & 786 & 16 \\
spaceman & 12{,}366 & 2{,}000 & 0 & 112 & 0 \\
sphere & 65{,}024 & 2{,}000 & 0 & 974 & 0 \\
thin shell & 32{,}256 & 2{,}000 & 0 & 322 & 0 \\
toy duck & 37{,}217 & 4{,}320 & 2{,}716 & 298 & 332 \\
\bottomrule
\end{tabular}
\end{table}

\begin{figure}[t]
    \centering
    \includegraphics[width=0.92\columnwidth]{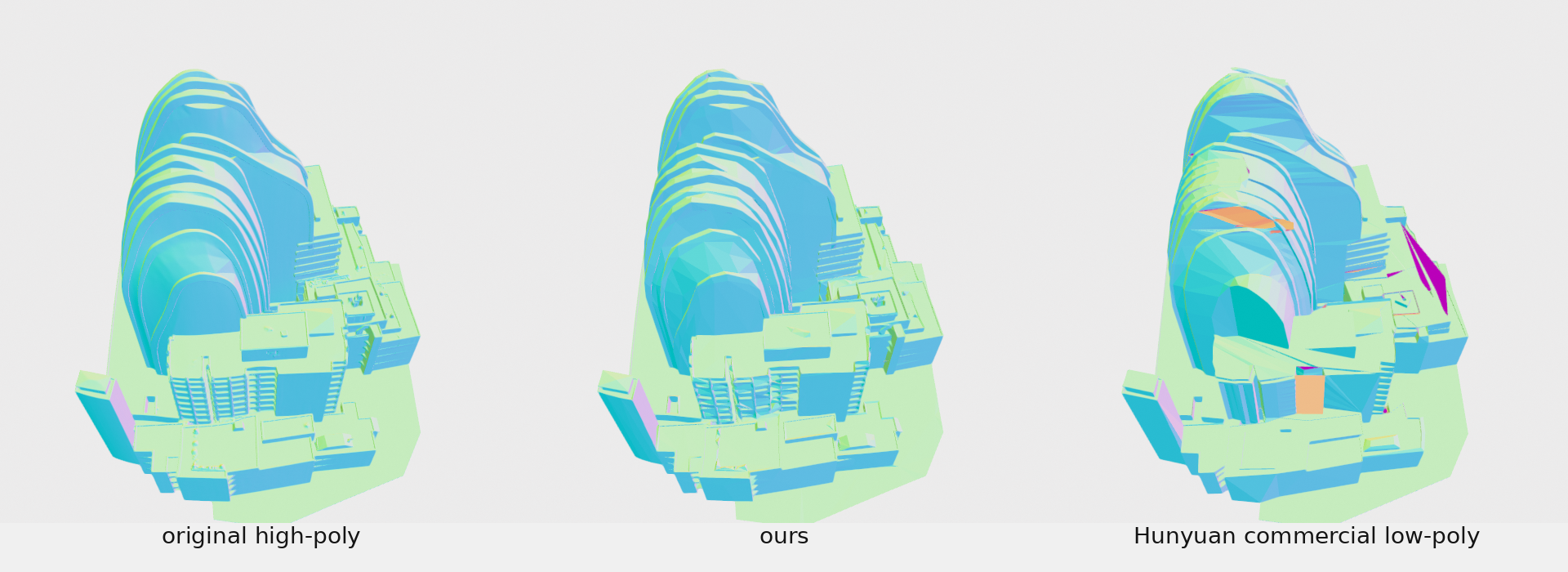}
    \caption{Detail inheritance on a training mesh (normal map): the original high-poly (left), our low-poly (center), and the Hunyuan commercial low-poly (right). The vendor output shows surface breakage on curved regions; ours preserves the curved fa\c{c}ade.}
    \label{fig:detail}
\end{figure}

\begin{figure*}[t]
    \centering
    \includegraphics[width=0.78\textwidth]{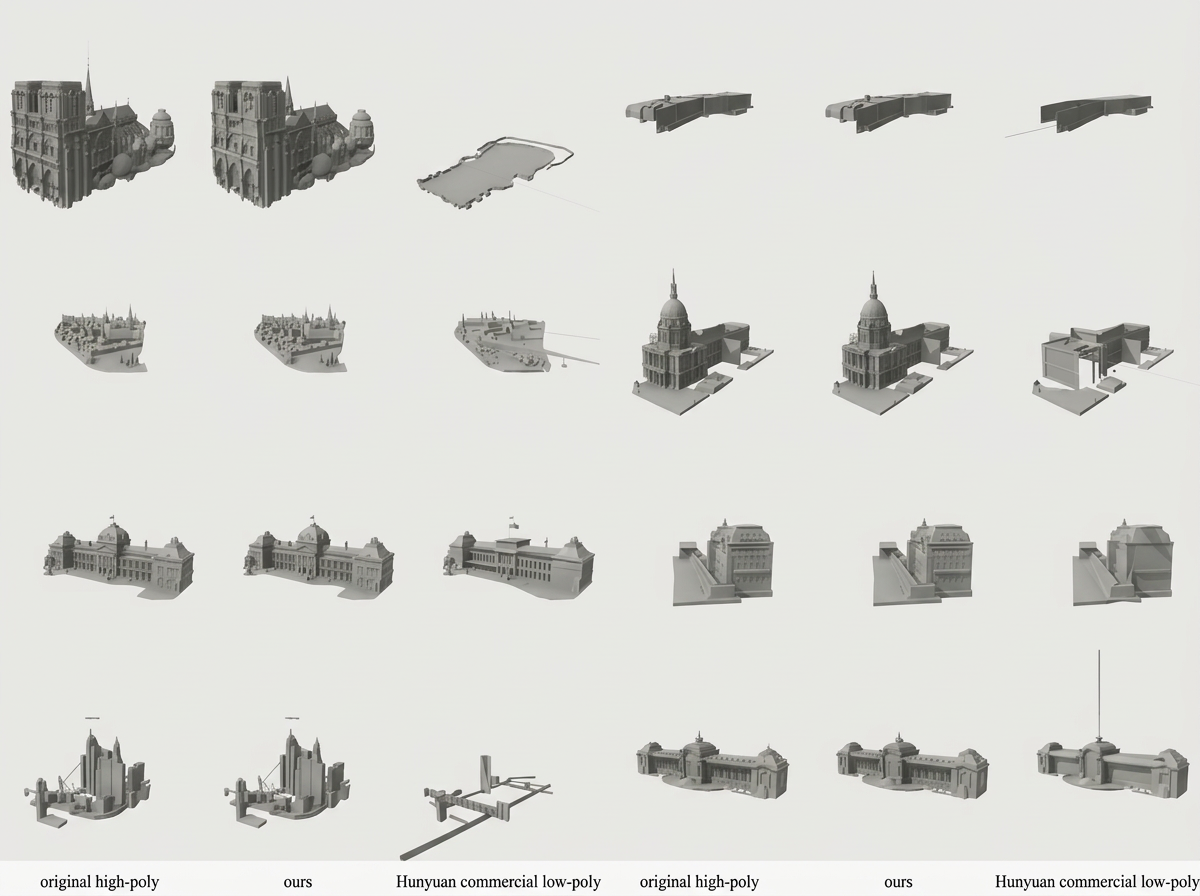}
    \caption{Qualitative comparison on eight training meshes (each triple: original high-poly, our low-poly, Hunyuan commercial low-poly). Our outputs preserve the full component structure of the 1.5M-face inputs; the vendor pipeline flattens catastrophically (top row), breaks interior walls, and grows spikes (bottom row)—the cliff failure modes.}
    \label{fig:gallery}
\end{figure*}

\subsection{Gate Statistics}\label{sec:exp-gate}

Each accept is a paired difference over the 19 validation meshes; Bernstein bounds (S5) put only the first two accepts above the resolution floor. Eq.~\eqref{eq:resolution} prices one paired comparison; under candidate multiplicity and validation reuse it is a resolution scale, not an anytime-valid certificate—the strict gate and the untouched test set carry the endpoint claim. Trajectory-level evidence escapes the per-step floor (all 19 meshes improve; supplementary S1b).

$\hat p = 5/21 \approx 0.24$ (95\% CI $[0.08, 0.47]$, Clopper--Pearson), $\hat\Delta \approx 0.02$, $\hat\sigma \approx 0.04$; regression candidates: 0 of 21 groups (95\% upper bound $0.143$ by the Rule of Three). The predicted 24-group gain $24\hat p\hat\Delta \approx 0.115$ matches the observed $0.145$.

\subsection{Component Ablation}\label{sec:exp-ablation}


Table~\ref{tab:ablation} reports the six single-factor arms (reflector, gate, retriever, selector) against the canonical configuration. The canonical run (same shuffle seed 44) is an \emph{independent rerun} of the main campaign—6/24 accepted here vs.\ 5/24 there; its higher endpoint reflects single-run seed variance. Every arm starts from $S_0 = 8.8358$ under the identical 24-group protocol and 19-mesh held-out gate.

\begin{table}[t]
\centering
\footnotesize
\caption{Single-factor ablation (seed 44).}
\label{tab:ablation}
\setlength{\tabcolsep}{3pt}
\begin{tabular}{llllccl}
\toprule
\textbf{Arm} & \textbf{Refl.} & \textbf{Gate} & \textbf{Retr.} & \textbf{Final} & $\boldsymbol\Delta$ & \textbf{Acc.} \\
\midrule
canonical & llm & strict\_gt & knn\_soft.& $8.9948$ & $+0.159$ & 6/24 \\
gate\_thresh. & llm & threshold & knn\_soft.& $8.9771$ & $+0.141$ & 2/24 \\
reflector\_rand. & rand. & strict\_gt & knn\_soft.& $8.9637$ & $+0.128$ & 2/24 \\
selection\_rand. & llm & strict\_gt$^{*}$ & knn\_soft.& $8.9593$ & $+0.124$ & 3/24 \\
retriever\_unif. & llm & strict\_gt & uniform & $8.9385$ & $+0.103$ & 1/24 \\
gate\_always & llm & always & knn\_soft.& $8.7501$ & $\mathbf{-0.086}$ & 17/24 \\
retriever\_none & llm & strict\_gt & none & $8.8358$ & $+0.000$ & 0/24 \\
\bottomrule
\multicolumn{7}{l}{\footnotesize $^{*}$selector replaced by uniform random choice; others submodular.}
\end{tabular}
\end{table}

Removing the gate is the only intervention that \emph{degrades} the final score ($-0.086$): 17/24 proposals are accepted, including negative-gain edits, and the population drifts below its start—without selection pressure, edit frequency is not learning. Thresholding ($\varepsilon{=}0.01$, $+0.141$) sits just under canonical. Random-attribute edits still improve (the gate filters noise) but trail LLM attribution by $0.031$; uniform retrieval ($+0.103$) and random selection ($+0.124$) fall below canonical—weighting and selection each contribute, neither load-bearing alone. Removing retrieval entirely stops evolution (all 24 groups inert). Across three shuffle seeds the canonical configuration ends at $8.983\pm0.011$—every seed above baseline, zero regressed updates in 72 groups. Attribution's $0.031$ edge exceeds this spread, and equal-budget search lands below the deployed state: random, Bayesian, and evolutionary search over sixteen clean--QEM configurations peak at $8.67$/$8.59$/$8.69$, a per-mesh oracle $8.75$, deployed $8.83$ (supplementary, S7).

Per-arm details are in the supplementary (S4); sensitivity orders: gating removal (the unique degradation, $-0.086$ vs.\ $S_0$), retrieval removal (no learning), then, against the canonical endpoint, weighting $\approx$ selection $\approx$ reflection.

\subsection{Comparison with Vendor Baseline}\label{sec:exp-vendor}

The commercial comparator is the Hunyuan3D vendor remesher (the low-poly output of the same pipeline that generated our training inputs), evaluated as released. Under the face-aligned verdict protocol (20 meshes scored at the vendor's own face counts; supplementary, S2b), our outputs win 5, tie 13, and lose 2. The protocols answer different questions: verdict ties mean both outputs meet the spec, whereas the discriminative comparison is the budget-aligned composite (17 wins to 3 losses, $p=0.001$).

Budget band: $1\%$ with $-50\%$ tolerance; per-mesh metrics for any caliber are in the supplementary (S2b).

A blind human study corroborates the verdict: the 24 training buildings are rendered as white-model blue-wireframe images, left/right randomized, and 9 independent participants (one session each, on a single workstation) pick the mesh that ``looks cleaner and more regular, with fewer tiny or stretched triangles.'' Aggregating 216 paired judgments: \textbf{ours 168, vendor 48, tie 0}; 8 of 9 subjects prefer ours (one prefers the vendor, 11--13); the one-sided sign test on the 8-to-1 split gives $p = 0.0195$ (Wilcoxon $p = 0.0039$). A test-set counterpart is not informative: the 20 unseen buildings are simple block-like geometry with visually indistinguishable outputs, so preference is reported on the training geometry where triangle quality is discriminable.

\subsection{Test-Set Evaluation}\label{sec:exp-testset}

On 20 unseen German city buildings (110k--800k faces, independently sourced; supplementary S2b), the trained skill keeps zero non-manifold edges on all 20 and consolidates each into a single dominant component (gate towers or residual dust aside, ${\le}195$ faces, against main bodies of 800--7{,}400), leading the composite (mean $8.82$ vs.\ $7.83$; paired $n{=}20$: $p = 0.001$, Wilcoxon $p = 0.0005$, 17 wins 3 losses; bootstrap 95\% CI of the mean gap $[0.51, 1.48]$). The vendor trails on planarity ($3.56$ vs.\ $3.22$), sharp edges ($5.79$ vs.\ $5.67$) despite $1.5\times$ our median faces, and worst-case distance (one-sided Hausdorff median $0.0105$ vs.\ $0.0081$); AutoRemesher fails on 45\% of inputs. These medians are the delivered approximation scale: the contract's $\varepsilon$ fixes the significance scale (Section~\ref{sec:contract}), not a per-output bound—at aggressive budgets the binding side is topology, which the outputs hold (zero non-manifold edges, single dominant component, all 20)—with full manifold metrics agreeing: self-intersections $1.5$ vs.\ $14.5$ median, degenerate faces $0/20$ vs.\ $12/20$, boundary edges $3/20$ vs.\ $13/20$ (supplementary, S2b). The zero-non-manifold transfer is direct evidence that ``clean before decimating'' generalizes (Figure~\ref{fig:gallery}).

Retrieval-only inference (KNN replay, no LLM calls) is zero-non-manifold on 19/20 of the same meshes and trails the full loop (20/20) but beats the vendor—transferable knowledge lives in the experience database itself. The evolution claim is tested directly on this untouched set: the zero-shot baseline $S_0$ (empty database, full loop; 16/20 buildings completed before a runner fault) reaches a median composite of $8.61$ against the trained skill's $8.85$—paired $n{=}16$, the trained skill wins 15 (Wilcoxon $p = 0.0001$)—while $S_0$ itself holds zero non-manifold edges on 15/16, locating the evolution gain on held-out data in quality beyond topology.

\subsection{Cross-Domain Transfer}\label{sec:exp-transfer}

We evaluate the same trained database (13 per-mesh records, all of Paris buildings) on nine non-architectural meshes (11k--250k faces; Table~\ref{tab:transfer}): retrieval replay, and a full zero-shot loop with an empty database. \textbf{(i)}~The clean-then-decimate lesson \emph{does} transfer: replay holds non-manifold ${\le}3$ edges on 7/9 meshes (148 spurious edges total outside one failure case), and the zero-shot loop stays at $\le 16$ edges on 7/9 meshes (bunny, 112, toy duck, 332, excepted). These are Stanford bunny and dragon, a reconstructed cup, a thin shell, a sphere, a toy duck, and three LATO.2 examples, spanning organic surfaces, thin walls, and up to ${\sim}1100$ components. \textbf{(ii)}~Numeric budget calibration does \emph{not} transfer: replayed absolute budgets pin outputs near the 2{,}000-face floor regardless of scale, whereas zero-shot respects the 1\% ratio on 6/9 (Figure~\ref{fig:transfer}a). \textbf{(iii)}~Negative transfer is real: on the thin-wall toy duck, building experience yields 2{,}716 non-manifold edges versus 332 without (Figure~\ref{fig:transfer}b). Procedural knowledge is domain-robust; numeric calibration is domain-bound.

\subsection{Tool-Tier Evolution}\label{sec:exp-operator}

The knowledge-tier experiments address what the agent \emph{knows}; this section evaluates the tool tier (\S\ref{sec:tier2}) through a pre-registered three-stage experiment. Six candidate operators—parameter variants beyond the built-in enumeration—are each evaluated on three probe meshes under pre-registered criteria (zero non-manifold edges, faces within $\pm20\%$, Chamfer within $3\times$ of built-in); all pass and are mounted (\S\ref{sec:tier2}).

On eight training meshes, every mounted operator competes under a pre-registered adoption rule (${\geq}15\%$ Chamfer gain, zero non-manifold edges): 0/8 adoptions in the primary run and 0/8 under two additional seeds (seeds 45 and 46); best per-mesh gain $0$--$3\%$ (median $1\%$). The uniform zero adoption—cross-confirming the equal-budget kernel A/B (supplementary, S2)—shows this parameter space is exhausted; a productive tool tier must mount structurally different operators.

In deployment, retrieval replay needs \emph{no} LLM call ($11.7$\,s per mesh on CPU, building test split); the full acquisition loop costs $4.3$/$12.2$\,min and the training campaign about a day (supplementary, S3). Under 12 single-dimension $\pm 50\%$ and 100 in-band weight perturbations the best accepted state beats the baseline in \emph{every} one ($18$--$19/19$ per mesh; $100/100$ positive); flipped micro-decisions are the marginal accepts near $\Delta_{\min} \approx 0.022$, per the resolution law.

%% file: sections/conclusion.tex
\section{Conclusion}\label{sec:conclusion}

We presented RemeshingAgent, the first agent for repair-based decimation of generated building meshes: a budget decomposition delimiting the cliff and mandating clean-first; a drift bound calibrated from the training trace; and a two-tier self-evolving agent under one strict gate—without a single regressed update. \emph{Limitations}: $S^\ast$ is a declared closure, not observed ground truth; ablation arms are single-run (canonical: three seeds, $8.983\pm0.011$, zero regressed updates); corpora are generated buildings, and numeric calibration does not transfer across domains (Section~\ref{sec:exp-transfer}); zero non-manifold edges and one dominant component are necessary, not sufficient, manifold evidence—outputs are not certified structural geometry. We intend to extend the evaluation to broader mesh families and to ground skill acquisition in synthetic validation so that new operators can be validated without human-curated corpora. Code, skills, database, prompts, thresholds, seeds, and splits will be released—a step toward agents that verify their own improvement.